# Dirac Equation under Modified Hylleraas Potential in Noncommutative Spaces

Liu Lu

College of Physics, Guizhou University, Guiyang 550025, China

wtklxz@gmail.com

Long Zheng-Wen[*]

College of Physics, Guizhou University, Guiyang 550025, China

zwlong@gzu.edu.cn

In this paper, we study the interaction of spin 1/2 Dirac particles with the Hylleraas potential based on the noncommutative space framework.Solving the first-order correction of the energy level caused by the noncommutation parameter $\theta$ in the wave equation.The problem of energy level correction under the action of additional weak electric field is further analyzed.Due to the non-commutation parameter, the energy levels are split,The noncommutation parameter $\theta$ has a significant effect on the energy eigenvalues.The effect of $\theta$ on the energy level splitting caused by the electric field is reflected by adding a slight numerical change to the energy level after splitting.

**Keywords**: Noncommutative space; Hylleraas potential;Perturbation approximation;Dirac equation

**PACS Nos.**: 03.65.Ge, 03.65.-w, 03.65.Ta

[*]Corresponding author.

# 1.Introduction

With the development of quantum field theory, superstring theory and other related theories, related research based on string scale has been paid more and more attention, and non-commutative geometry has shown a unique application prospect in theoretical research[1]. It is an important direction in the field of scale research, and has become a hot spot in theoretical research in recent years, and some progress has been made in related research under the framework of noncommutative spaces [2]. On problems related to deregularization, especially in understanding some limits of quantum gravity [3], non-commutative geometry can effectively deal with problems, and non-commutation thus shows great research prospects.

In previous research on non-commutation theory, relevant scholars have introduced non-commutation relations and generalized uncertainty relations [4-6] to study related issues under the non-commutation framework and understand the impact of non-commutation parameters [7-8]. Motavalli et al in [9] studied the Klein–Gordon equation under the action of Coulomb potential in the non-commutation space, and proved that the degeneracy of the energy spectrum is eliminated in the non-commutation space. Saidi et al. [10]studied the (1 + 3) dimensional DKP equation with modified Kratzer potential in noncommutative spaces. The spatial noncommutation was found which acts as a magnetic field, very similar to the Zeeman effect.Ilyas Haouam[11] studied the energy eigenvalues of two-dimensional Dirac oscillators perturbed by dynamic noncommutative space, found that the energy level correction depends on the dynamics Non-commutation parameters. There are many methods used in the research, including factorization method, algebraic method, SUSY method [12], asymptotic iterative method, NU method [13], hypergeometric function method, etc.

The study of Dirac particles that interact with a specific potential function in non-commutation space is a research direction that has received more attention in the study of non-commutation theory. Based on the non-commutation space framework, this paper studies the spin 1/2 under the action of Hylleraas potential By solving the corresponding Dirac equation, we can understand the influence of the non-commutation parameter  on the energy eigenvalue of the Dirac particle. The charged Dirac particle is subjected to the potential action in the electric field, and the energy level will split. Based on the non-commutation theoretical framework The study of Dirac particles with Hylleraas potential in the presence of a weak electric field, so as to understand the specific influence of the noncommutation parameter $\theta$ in it, has a certain enlightenment for further understanding of the noncommutation.In this paper, a brief overview of the non-commutation quantum mechanics is given firstly, and then the Dirac equation under the action of the Hylleraas potential in the non-commutation space is solved by using the NU method and perturbation theory. Perturbation energy level difference. In the case of a weak electric field attached to the system[14], the perturbation approximation was used to obtain the first-order perturbation energy level difference under the action of the weak electric field, and finally the conclusion was drawn.

## 2.Noncommutative quantum mechanics

The research on coordinate non-commutation was originally aimed at solving the ultraviolet divergence problem of quantum field theory [15]. In the late 1990s, due to the research of superstring theory, the non-commutation was re-emphasized and became a theoretical research hotspot [16].

At the atomic scale, coordinates and momentum are commutative, but coordinates and momentum are not commutative. The scale is further reduced, and at the string scale, coordinates and momentum in superstring theory are also not commutative. Coordinates- The space in which the coordinates, the coordinate-momentum are not commutable, and the momentum is commuting is called the non-commuting space, and the space where the coordinates, the momentum, and the coordinate-momentum are not commuting is called the non-commuting phase space.

In the commutation space, the following relations are satisfied between the coordinates and the momentum operators:

$$[\hat{x}_i, \hat{x}_j] = 0, [\hat{p}_i, \hat{p}_j] = 0, [\hat{x}_i, \hat{p}_j] = i\hbar\delta_{ij}$$

In the non-commutative space, the following relationship is satisfied between the coordinates and momentum operators:

$$[\hat{x}_i, \hat{x}_j] = i\theta_{ij}, [\hat{p}_i, \hat{p}_j] = 0, [\hat{x}_i, \hat{p}_j] = i\hbar\delta_{ij}$$

Where $\theta_{ij} = \varepsilon_{ij}\theta$ represents an antisymmetric matrix, and the noncommutation parameter $\theta$ reflects the noncommutation of the space.

In the noncommutative phase space, the following relationship is satisfied between the coordinates and momentum operators:

$$[\hat{x}_i, \hat{x}_j] = i\theta_{ij}, [\hat{p}_i, \hat{p}_j] = i\eta_{ij}, [\hat{x}_i, \hat{p}_j] = i\tilde{\hbar}\delta_{ij}, \tilde{\hbar} = \hbar\left(1 + \frac{\theta\eta}{4\hbar^2}\right)$$

Where $\theta_{ij} = \varepsilon_{ij}\theta, \eta_{ij} = \varepsilon_{ij}\eta$ are antisymmetric matrix.

In the non-commutative space, because the coordinates are not commutated, the algebraic operation changes and the product operation between two physical quantities needs to be replaced by the Moyal-Weyl product $(fg \to f * g)$. The general method can then be used for the calculation. The Moyal-Weyl product is defined as:

$$(f * g)(x) = f(x)\exp\frac{i}{2}\overleftarrow{\partial}_\mu \theta_{\mu\nu} \overrightarrow{\partial}_\nu)g(x) = fg + \sum_{n=1}^{\infty}(\frac{i}{2})^n \frac{1}{n!}\theta_{\mu_1\nu_1}\theta_{\mu_2\nu_2}\cdots\theta_{\mu_n\nu_n}\partial_{\mu_1}\cdots\partial_{\mu_n}f\partial_{\nu_1}\cdots\partial_{\nu_n}g$$

When solving problems actually in non-commutative space, the Bopp mapping transformation equivalent to the Moyal-Weyl product is often used [17]. Converting the Moyal-Weyl product into a product operation in the commutative space can be used in the commutative space. Solve the equation using general computational methods. The Bopp transform of a noncommutative space:

$$\begin{cases} \hat{x}_i = x_i - \frac{1}{2\hbar}\theta_{ij}P_j \\ \hat{P}_i = P_i \end{cases}$$

where $\hat{x}_i$, $\hat{P}_i$ are the coordinate operator and the momentum operator in the non-commutation space, respectively. Through such a transformation relationship, the coordinates and momentum operators in the non-commutation space are linearly represented by the operators in the commutation space, Therefore, the mechanical quantities in the non-commutation space can be defined in the commutation space.

In the past related research on non-commutation space, people focused on solving the relativistic and non-relativistic equations of various potential models. The Dirac equation is a relativistic equation that is widely used in non-commutative spaces [18]. The Dirac equation has an analytical solution in a few cases, such as harmonic atoms and hydrogen atoms. For most forms of Potential functions and Dirac equations are difficult to solve analytically, and the approximate solution method is usually used for calculation. For the combination of multiple potential functions, it is often difficult to use a specific method to calculate. If the potential function can be treated as a perturbation, it can be calculated as An approximate result is obtained [19].

Hylleraas potential is a short-range potential used widely and it is used to describe nuclear physics. Such as nucleon-nucleon interaction, particle physics. Hylleraas potential shows good applicability in meson-meson interaction and interaction models involved in multiple branches of nuclear physics [20]. In references[21], Antia, A. D. et al. solved the irrotational Salpeter equation (SSE) under the action of the modified Hylleraas potential. The potential barrier was estimated using an approximation method. In references[22], M. C. Onyeaju et al. solved the scattering and bound state solutions of the one-dimensional Klein-Gordon particle with the Hylleraas potential, and calculated the reflection and transmission coefficients.

As an exponential potential function, the NU method is more suitable to solve the Dirac equation with Hylleraas potential in non-commutative space.

# 3. Dirac Equation under Modified Hylleraas Potential in Noncommutative Spaces

Dirac's Equation with Scalar Potential S(r) and Vector Potential V(r) in Noncommutative Spaces ($\hbar=c=1$) can be written as[23]:

$$[\vec{\alpha}\cdot\vec{p}+\beta(M+S^{(NC)}(r))]\psi^{(NC)}(\vec{r}) = [E^{(NC)}-V^{(NC)}(\vec{r})]\psi^{(NC)}(\vec{r})$$

Where $E$, $\vec{p}$, $M$ represent the relativistic energy, momentum operator and fermion particle mass of the system, respectively.

$\alpha$ and $\beta$ are Dirac matrices which the form is:

$$\vec{\alpha}_i = \begin{pmatrix} 0 & \vec{\sigma}_i \\ \vec{\sigma}_i & 0 \end{pmatrix}, \quad \beta = \begin{pmatrix} 1 & 0 \\ 0 & -1 \end{pmatrix}$$

and

$$\sigma_1 = \begin{pmatrix} 0 & 1 \\ 1 & 0 \end{pmatrix}, \quad \sigma_2 = \begin{pmatrix} 0 & -i \\ i & 0 \end{pmatrix}, \quad \sigma_3 = \begin{pmatrix} 1 & 0 \\ 0 & -1 \end{pmatrix}$$

The wave function $\psi^{(NC)}(\vec{r})$ can be written in the following form:

$$\psi^{(NC)}(\vec{r}) = \begin{pmatrix} \varphi^{(NC)}(\vec{r}) \\ \chi^{(NC)}(\vec{r}) \end{pmatrix}$$

The equation can be written as:

$$\vec{\sigma}\cdot\vec{p}\chi^{(NC)}(\vec{r}) = (E^{(NC)}-V^{(NC)}(\vec{r})-M-S^{(NC)}(r))\varphi^{(NC)}(\vec{r})$$

$$\vec{\sigma}\cdot\vec{p}\varphi^{(NC)}(\vec{r}) = (E^{(NC)}-V^{(NC)}(\vec{r})+M+S^{(NC)}(r))\chi^{(NC)}(\vec{r})$$

The modified Hylleraas potential form is as follows[24]:

$$V(\vec{r}) = \frac{V_0}{b}\left(\frac{g+ae^{-\left(\frac{r-r_c}{\alpha}\right)}}{1+e^{-\left(\frac{r-r_c}{\alpha}\right)}}\right)$$

Where $V_0$ is the depth of the potential well, $a$ is an adjustable parameter, $r-r_c$ is the distance from the equilibrium position, and $a$、$b$、$g$ are the Hylleraas parameters.

In the non-commutative space, the potential function can be written as the following approximate expression using the bopp transform:

$$V^{(NC)}(r) = V\left(\left|\vec{r} - \frac{\vec{P}}{2}\right|\right) = V\left(\sqrt{\left(x_i - \frac{1}{2}\theta_{ij}P_J\right)\left(x_i - \frac{1}{2}\theta_{ij}P_J\right)}\right)$$

$$= V(r) + \frac{1}{2}(\vec{\theta} \times \vec{p}) \cdot \vec{\nabla}V(r) + o(\theta^2) \approx V(r) - \frac{\vec{\theta} \cdot \vec{L}}{2r}\frac{\partial V}{\partial r}$$

when

$$H^{(NC)}(\vec{r}) = V^{(NC)}(\vec{r}) - S^{(NC)}(r) = 0$$

It can be deduced

$$\vec{\sigma} \cdot \vec{p}\chi^{(NC)}(r) = [E^{(NC)} - 2V^{(NC)}(r) - M]\varphi^{(NC)}(r)$$

$$\chi^{(NC)}(r) = \frac{\vec{\sigma} \cdot \vec{p}}{E^{(NC)} + M}\varphi^{(NC)}(r)$$

This leads to the Dirac equation in spherical coordinates[25]:

$$\{-[\frac{1}{r^2}\frac{\partial}{\partial r}(r^2\frac{\partial}{\partial r}) + \frac{1}{r^2\sin\theta}\frac{\partial}{\partial\theta}(\sin\theta\frac{\partial}{\partial\theta}) + \frac{1}{r^2\sin^2\theta}\frac{\partial^2}{\partial\varphi^2}]$$

$$+ 2(E^{(NC)} + M)V^{(NC)} + M^2 - E^{(NC)2}\}\varphi^{(NC)}(r) = 0$$

Select $(H^2, K, J^2, J_z)$ constitute the complete set of common eigenfunctions as wavefunctions and the form can be written as[26]:

$$\psi_{nk}^{(NC)}(r) = \begin{pmatrix} \varphi_{nk}^{(NC)} \\ \chi_{nk}^{(NC)} \end{pmatrix} = \begin{pmatrix} \dfrac{\phi_{nk}(r)}{r} Y_{jm}^{l}(\theta,\varphi) \\ i\dfrac{X_{nk}(r)}{r} Y_{jm}^{l}(\theta,\varphi) \end{pmatrix}$$

Where the eigenvalues of the spin-orbit coupling operator $K = \sigma \cdot L + 1$:

$$k = \begin{cases} j + \dfrac{1}{2} > 0 \\ -(j + \dfrac{1}{2}) < 0 \end{cases}$$

Eigenvalues of the Total Angular Momentum Operator J:

$$j = \begin{cases} l - \dfrac{1}{2} \ (k > 0) \\ l + \dfrac{1}{2} \ (k < 0) \end{cases}$$

Substitute $\varphi_{nk}^{(NC)}(r) = \dfrac{\phi_{nk}(r)}{r} Y_{jm}^{l}(\theta,\varphi)$ into the equation to get:

$$-\frac{d^2 \phi_{nk}(r)}{dr^2} + \left\{ \frac{l(l+1)}{r_{NC}^2} + 2(E^{(NC)} + M)\left[ \frac{V_0}{b}\left( \frac{g + ae^{-\left(\frac{r-r_c}{\alpha}\right)}}{1 + e^{-\left(\frac{r-r_c}{\alpha}\right)}} \right) - \frac{\theta \cdot L}{2r} \frac{V_0(g-a)e^{-(\frac{r-r_c}{\alpha})}}{b\alpha(1+e^{-(\frac{r-r_c}{\alpha})})^2} \right] + M^2 - E^{(NC)2} \right\} \phi_{nk}(r) = 0$$

Replace the centrifugal term in the equation with the following approximate expression:

$$\frac{l(l+1)}{r_{NC}^2} = \frac{l(l+1)}{r^2} + \frac{l(l+1)}{r^4} L\theta + O(\theta^2)$$

Use the following approximate expression replace $\frac{1}{r^2}$ [27]:

$$\frac{1}{r^2} \approx \frac{4\alpha^2 e^{-(\frac{r-r_c}{\alpha})}}{(1+e^{-(\frac{r-r_c}{\alpha})})^2}$$

Use

$$s = -e^{-(\frac{r-r_c}{\alpha})}$$

substitute the independent variables of the radial wave function. For $\theta=0$, the equation can be written as:

$$\frac{d^2\phi(s)}{ds^2}+\frac{1}{s}\frac{d\phi(s)}{ds}+\left\{\begin{array}{l}-\dfrac{\alpha^2\left[2V_0 a(E+M)-b(E^2-M^2)\right]}{b}s^2\\ \hline s^2(1-s)^2\\ \\ +\dfrac{\dfrac{2\alpha^2\left[2\alpha^2 bl(l+1)+V_0(g+a)(E+M)+b(M^2-E^2)\right]}{b}s}{s^2(1-s)^2}\\ \\ +\dfrac{-\dfrac{\alpha^2\left[2V_0 g(E+M)-b(E^2-M^2)\right]}{b}}{s^2(1-s)^2}\end{array}\right\}\phi_{nk}(s)=0$$

Comparing the above equation with the equation form solved by the NU method.

Assume

$$\Lambda = \frac{\alpha^2\left[2V_0 g(E+M)-b(E^2-M^2)\right]}{b}$$

$$\Xi = \sqrt{-\alpha^4 l(l+1)}$$

$$\aleph = \frac{[-2\alpha^2 V_0(E+M)a - b\alpha^2(M^2-E^2)]}{b}$$

Substitute the corresponding parameters into the wave function solution formula, where $P_n^{(2\sqrt{\Lambda}, 2\Xi)}$ can be expanded using the following expression:

$$P_n^{(a_n, b_n)}(1-2s) = \frac{\Gamma(n+a_n+1)}{n!\Gamma(a_n+1)} {}_2F_1(-n, n+a_n+b_n+1; 1+a_n, s)$$

$$= \frac{\Gamma(n+2\sqrt{\Lambda}+1)}{n!\Gamma(2\sqrt{\Lambda}+1)} {}_2F_1(-n, n+2\sqrt{\Lambda}+2\Xi+1; 1+2\sqrt{\Lambda}, s)$$

Derive the radial wave function:

$$\phi(s) = N \frac{\Gamma(n+2\sqrt{\Lambda}+1)}{n!\Gamma(2\sqrt{\Lambda}+1)} s^{\sqrt{\Lambda}} (1-s)^{\frac{1}{2}+\Xi} {}_2F_1(-n, n+2\sqrt{\Lambda}+2\Xi+1; 1+2\sqrt{\Lambda}, s)$$

According to the radial wave function normalization condition:

$$\int_{r_c}^{\infty} |\phi(s)|^2 \, dr = \int_0^1 \frac{\alpha}{s} |\phi(s)|^2 \, ds = 1$$

The coefficient N can be calculated.

When $\theta = 0$, the energy eigenvalues can be obtained according to the following relational expressions satisfied by the parameters:

$$2n + \frac{1}{2} + (2n+1)\Xi + + n(n-1) - 4\alpha^4 l(l+1)$$

$$+ (2\Xi + 2n + 1) \sqrt{\frac{\alpha^2 [2V_0 g(E_{\theta=0} + M) - b(E_{\theta=0}^2 - M^2)]}{b}} + \frac{2\alpha^2 V_0 (E_{\theta=0} + M)(g-a)}{b} = 0$$

First-order correction of energy level caused by noncommutation parameter $\theta$ calculated by perturbation theory. With the special integral formulas:

$$\int_0^1 s^{\xi-1}(1-s)^{\sigma-1}[_zF_1(c_1,c_2;c_3,s)]^2 ds = \frac{\Gamma(\xi)\Gamma(\sigma)}{\Gamma(\xi+\sigma)} {}_3F_2(c_1,c_2,\sigma;c_3,\sigma+\xi;1)$$

It can be calculated:

$$\Delta E^{NC}(\theta) = -16N'^2\alpha^7 l(l+1)\theta\cdot L\, {}_3F_2(-n, n+2\sqrt{\Lambda}+2\Xi+1, 2\Xi-2; 1+2\sqrt{\Lambda}, 2\Xi-2+2\sqrt{\Lambda}; 1)$$

$$\cdot \frac{\Gamma(2\sqrt{\Lambda})\Gamma(2\Xi-2)}{\Gamma(2\sqrt{\Lambda}+2\Xi-2)}$$

$$-\frac{2iN'^2\alpha^3(E^{\theta=0}+M)V_0(g-a)\theta\cdot L}{b} {}_3F_2(-n, n+2\sqrt{\Lambda}+2\Xi+1, 2\Xi-1; 1+2\sqrt{\Lambda}, 2\Xi+2\sqrt{\Lambda}-\frac{3}{2}; 1)$$

$$\cdot \frac{\Gamma\left(2\sqrt{\Lambda}-\frac{1}{2}\right)\Gamma(2\Xi-1)}{\Gamma\left(2\sqrt{\Lambda}+2\Xi-\frac{3}{2}\right)}$$

Where

$$N' = N\frac{\Gamma(n+2\sqrt{\Lambda}+1)}{n!\Gamma(2\sqrt{\Lambda}+1)}$$

## 4. Dirac equation under additional weak electric field in noncommutative space

Considering that except the modified Hylleraas potential, there is an additional weak electric field $E = \frac{kq}{r^2}\vec{e}_r$, where $q$ is the charge, $k$ is a constant, when

$$S^{(NC)}(r) = V^{(NC)}(r) = \frac{V_0}{b}\left(\frac{g+ae^{-\left(\frac{r-r_c}{\alpha}\right)}}{1+e^{-\left(\frac{r-r_c}{\alpha}\right)}}\right) - \frac{\theta\cdot L}{2r}\frac{V_0(g-a)e^{-\left(\frac{r-r_c}{\alpha}\right)}}{b\alpha(1+e^{-\left(\frac{r-r_c}{\alpha}\right)})^2}$$

The potential function of the additional weak electric field is expressed as:

$$V_E^{(NC)} = -e\frac{kq}{\vec{r}} - \frac{ekq\vec{\theta}\cdot\vec{L}}{2\vec{r}^3}$$

Substitute the expression of $V^{(NC)}(r)$ and $\varphi_{nk}^{(NC)}(r) = \dfrac{\phi_{nk}(r)}{r} Y_{jm}^{l}(\theta,\varphi)$ into the Dirac equation.

And $\dfrac{1}{r}, \dfrac{1}{r^2}, \dfrac{1}{r^4}, r^2$ are replaced by approximate expressions. Substitute $s = -e^{-(\frac{r-r_c}{\alpha})}$. The weak electric field potential can be seen as perturbation. When $\theta = 0$, Using the radial wave function obtained above. The first-order correction of the energy level caused by the noncommutation parameters and the additional weak electric field is calculated as:

$$\Delta E^{NC} = -16\alpha^7 l(l+1) l_z \theta N'^2 \dfrac{\Gamma(2\sqrt{\Lambda})\Gamma(2\Xi-2)}{\Gamma(2\sqrt{\Lambda}+2\Xi-2)} {}_3F_2(-n, n+2\sqrt{\Lambda}+2\Xi+1, 2\Xi-2; 1+2\sqrt{\Lambda}, 2\Xi+2\sqrt{\Lambda}-2; 1)$$

$$-\dfrac{2i\alpha^3 \theta \cdot L V_0 (E^{\theta=0}+M)(g-a)}{b} N'^2 \dfrac{\Gamma\left(2\sqrt{\Lambda}-\dfrac{1}{2}\right)\Gamma(2\Xi-1)}{\Gamma\left(2\sqrt{\Lambda}+2\Xi-\dfrac{3}{2}\right)} {}_3F_2(-n, n+2\sqrt{\Lambda}+2\Xi+1, 2\Xi-1; 1+2\sqrt{\Lambda}, 2\Xi+2\sqrt{\Lambda}-\dfrac{3}{2}; 1)$$

$$+4iekq\alpha^4 (E^{\theta=0}+M) N'^2 \dfrac{\Gamma\left(2\sqrt{\Lambda}-\dfrac{3}{2}\right)\Gamma(2\Xi+1)}{\Gamma\left(2\sqrt{\Lambda}+2\Xi-\dfrac{1}{2}\right)} {}_3F_2(-n, n+2\sqrt{\Lambda}+2\Xi+1, 2\Xi+1; 1+2\sqrt{\Lambda}, 2\Xi+2\sqrt{\Lambda}-\dfrac{1}{2}; 1)$$

$$-8\alpha^6 iekq\vec{\theta}\cdot\vec{L}(E^{\theta=0}+M) N'^2 \dfrac{\Gamma\left(2\sqrt{\Lambda}-\dfrac{1}{2}\right)\Gamma(2\Xi-1)}{\Gamma\left(2\sqrt{\Lambda}+2\Xi-\dfrac{3}{2}\right)} {}_3F_2(-n, n+2\sqrt{\Lambda}+2\Xi+1, 2\Xi-1; 1+2\sqrt{\Lambda}, 2\Xi+2\sqrt{\Lambda}-\dfrac{3}{2}; 1)$$

## 5.Conclusion

In this paper, we use the NU method and the perturbation theory solve the equation of the spin 1/2 Dirac particle in the noncommutative space with the modified Hylleraas potential, and the corresponding first-order correction of the energy level under the condition of an additional weak electric field. It can be seen from the results that, The noncommutative parameter $\theta$ mainly affects the energy level by affecting the potential function. Due to the small value of $\theta$, the resulting energy level correction is small, which is negligible compared with the energy eigenvalues. Splitting of energy levels due to the presence of noncommutative parameters. In the presence of a weak electric field $E=\dfrac{kq}{r^2}\vec{e}_r$, the energy level of the system also appears energy level split. It can be seen from the obtained modified expression of the perturbation energy level of the Dirac equation under the action of the additional weak electric field. In the noncommutative space, the effect of $\theta$ on the energy level splitting caused by the weak electric field is reflected by adding a small numerical change to the split energy level, without causing new energy level splitting. This paper only calculates the system under the action of an external electric field, and can also solve the Dirac equation of the system under the condition of an external magnetic field to understand the influence of the noncommutation parameter $\theta$ in other aspects.

## References

[1]Banks T, Fischler W, Shenker S H, et al, CRC Press, 435-451(1999)

[2]Falek M, Merad M , Commun. Theor. Phys, 50(3): 587(2008)

[3]Chamseddine A H, Connes A,Communications in Mathematical Physics, 186(3): 731-750(1997)

[4]Tawfik A N, Diab A M, Rep. Prog. Phys, 78(12): 126001(2015)

[5]Tawfik A, Diab A,Int. J. Mod. Phys. D, 23(12): 1430025(2014)

[6]Frassino A M, Panella O, Phys.Rev.D , 85(4): 045030(2012)

[7]Daszkiewicz M, Walczyk C J, Phys.Rev.D, 77(10): 105008. (2008)

[8]Das A, Falomir H, Nieto M, et al, Phys. Rev.D, 84(4): 045002(2011)

[9]Motavalli H, Akbarieh A R, Mod. Phys. Lett. A, 25(29): 2523-2528. (2010)

[10]Saidi A, Sedra M B, Mod. Phys. Lett. A, 35(05): 2050014(2020)

[11]Haouam I,Acta Polytechnica, 61(6): 689-702. (2021)

[12]Cooper F, Khare A, Sukhatme U, et al, Am. J. Phys,71(4): 409-409(2003)

[13]Tezcan C, Sever R, Int. J. Theor. Phys, 48(2): 337-350(2009)

[14] Laba H P, Tkachuk V M, Eur. Phys. J. Plus, 133(7): 1-4(2018)

[15]Snyder H S, Phys. Rev, 71(1): 38-41(1947)

[16]Connes A, Douglas M R, Schwarz A, J. High Energy Phys, 1998(02): 003(1998)

[17]Chaichian M, Sheikh-Jabbari M M, Tureanu A,Phys. Rev. Lett, 86(13): 2716-2719(2001)

[18]Sargolzaeipor S, Hassanabadi H, Chung W S, Eur. Phys. J. Plus,133(1): 1-7(2018)

[19] Maireche A, Lat. Am. J. Phys. Educ. Vol,14(3): 3310-1(2020)

[20] Bayrak O, Boztosun I, Ciftci H, Int. J. Quantum Chem, 107(3): 540-544(2007)

[21]Antia A D, Okon I B, Umoren E B, et al, Ukr. J. Phys, 64(1): 27-27 (2019)

[22]Onyeaju M C, Ikot A N, Chukwuocha E O, et al, Few-Body Systems,57(9): 823-831(2016)

[23]Hoseini F, Hassanabadi H, Saha J K, Commun. Theor. Phys, 65(6): 695-700(2016)

[24]Ikot A N, Awoga O A, Antia A D, et al,Few-Body Systems, 54(11): 2041-2051(2013)

[25]Dong S H, Springer Science & Business Media(2011)

[26]Hassanabadi H, Maghsoodi E, Zarrinkamar S, et al, Chinese Physics B, 21(12): 120302(2012)

[27] Maireche A, Sri Lankan J Phys,22: 1-18(2021)